\documentclass[
reprint,
superscriptaddress,
nofootinbib,
amsmath,
amssymb,
aps,
prc,
floatfix,
]{revtex4-2}
\usepackage{latexsym}
\usepackage{xcolor}
\newcommand{\bqa}{\begin{eqnarray}}
\newcommand{\eqa}{\end{eqnarray}}
\usepackage{amsmath}
\usepackage{array}
\usepackage{graphicx}
\graphicspath{{./}} 
\usepackage{dcolumn}
\usepackage{bm}
\usepackage[utf8x]{inputenc}
\usepackage{hyperref}
\hypersetup{linkcolor=blue}
\begin{document}
\preprint{APS/123-QED}
\title{Probing the onset of hydrodynamization in peripheral p-Pb collisions at $\sqrt{s_{NN}} =$ 5.02 TeV }
\author{Nikhil Hatwar}
\email{nikhil.hatwar@gmail.com}
\affiliation{Department of Applied Sciences and Humanities, Physics section, COEP Technological University, Pune 411005, India}
\affiliation{Department of Physics, Indian Institute of Technology Bombay, Mumbai 400076, India}

\author{Sadhana Dash}
\affiliation{Department of Physics, Indian Institute of Technology Bombay, Mumbai 400076, India}
\author{Basanta Kumar Nandi}
\affiliation{Department of Physics, Indian Institute of Technology Bombay, Mumbai 400076, India}
\date{\today}%
\begin{abstract}
	
We estimate the multiplicity scale below which the hydrodynamic description of small systems, such as p-Pb collisions, becomes increasingly sensitive to non-hydrodynamic effects. 
The sensitivity of experimental observables, namely the elliptic flow coefficient, is investigated by varying the shear relaxation time within second-order relativistic viscous hydrodynamics.
To achieve this, a representative system of peripheral p-Pb collisions at $\sqrt{s_{NN}} =$ 5.02 TeV is simulated utilizing the state-of-the-art JETSCAPE event generator. 
The soft-sector dynamics are modeled via a sequence of initial conditions, a pre-equilibrium stage, hydrodynamics, and a hadronic afterburner. 
Transverse momentum spectra and rapidity distributions for light-flavored hadrons are extracted and compared with experimental data. The resulting increase in elliptic flow fluctuations indicates a breakdown of fluid behavior at a rapidity density of $dN/dy \sim 10\text{--}15$ for p-Pb collisions at $\sqrt{s_{NN}} =$ 5.02 TeV.

\end{abstract}
\maketitle
\section{\label{sec:intro} Introduction}
Over the last few decades, we have slowly gathered mounting indirect evidence of the production of deconfined quark-gluon matter in relativistic heavy-ion collisions carried out at BNL's Relativistic Heavy Ion Collider and CERN's Large Hadron Collider~\citep{Foka2016Nov_HIC1,Foka2016Nov_HIC2,PHENIXwhitepaper,STARwhitepaper,PHOBOSwhitepaper,BRAHMSwhitepaper}. The focus is now shifting toward finding the properties of this phase of strongly interacting matter. However, the discovery of most of these quark--gluon plasma (QGP) signatures in small systems like proton-nucleus and high-multiplicity proton-proton collisions has been puzzling~\citep{pp_ridge_CMS, CMSpPbRidge2013, Nagle2018Oct_smallsys1,Dusling_2016, BibEntry2017Jun_smallsys2}. 

\begin{figure}
	\centering
	\includegraphics[scale=0.25]{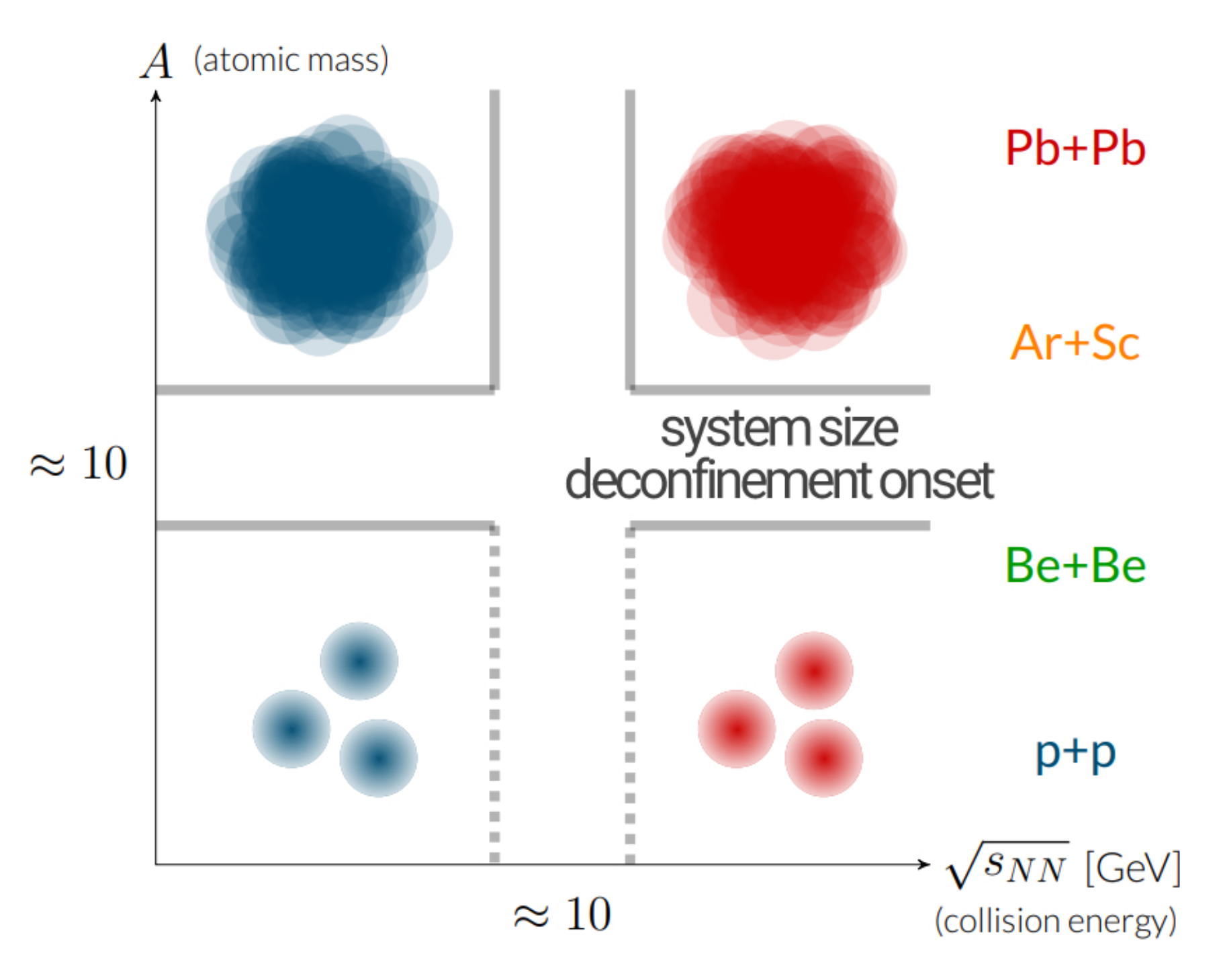}
	\caption{Cartoon of collision domains for varying atomic mass and collisional energy. The three blue circles and the large blue blob at low collisional energy represent cold nuclear matter. The large red blob represents the QGP formation regime. The three small red circles represent a region where the nature of the QCD matter produced is debatable.
	Reproduced from~\citep{Turko2018,Aduszkiewicz2018Mar}. \href{https://creativecommons.org/licenses/by/4.0/}{CC BY 4.0}.}
	\label{fig:collision_regimes}
\end{figure}

Currently, we are confronted with the question of whether these signatures are a result of cold nuclear matter effects~\citep{Armesto2018}, genuine QGP-like hot nuclear matter effects~\citep{Rafelski2015Sep_QGPsignatures}, or something in-between~\citep{Strickland2019Feb}.
Figure~\ref{fig:collision_regimes} shows a cartoon of collision domains for varying nuclear mass (which represents the system size of the collision) and collisional energy. In the domain of large collisional energy and large system size, we have convincing evidence of QGP formation, shown as a large red blob. At low collisional energy, we have nucleons and/or partons scattering off each other, represented by the blue color. There is no consensus about the nature of the medium produced for smaller system sizes and large collisional energy, represented by the three red circles in figure~\ref{fig:collision_regimes}. There should exist a threshold value of collisional energy, for a suitably large collision system, beyond which hot QCD matter could form~\citep{energy_deconfine1,energy_deconfine2}. If there is no formation of a deconfined medium in small systems like proton-proton collisions at large energies, there should exist another threshold of system size for which hot QCD matter could form. In the present study, we are trying to estimate this system size threshold, marked as the \textit{system size deconfinement onset} in 
figure~\ref{fig:collision_regimes}, for peripheral p-Pb collisions at $\sqrt{s_{NN}} =$ 5.02 TeV.

Relativistic hydrodynamics has emerged as a key effective theory for modeling the QGP phase of high-energy collisions~\citep{Gale2013Apr_hydro,HeinzSnellings2013}. The early phenomenological success of the hydrodynamic description of collective expansion in heavy-ion collisions indicated the formation of locally equilibrated isotropic matter as early as 1 fm/\textit{c}~\citep{Heinz2004_fastTherm}. It was challenging to explain such a rapid transition from a strong color-field configuration to a locally thermalized system of quarks and gluons~\citep{bottomUp,BergesHellerMazeliauskasVenugopalan2021}. It was found that, starting from various initial conditions, the evolution of pressure anisotropy is accurately described by the hydrodynamic gradient expansion. This surprising result indicated that the phenomenological success of hydrodynamics does not necessarily require early thermalization or even isotropization in heavy-ion collisions~\citep{attractor1,Spalinski2018Jan,2021AttractorReview}. Hence, a more appropriate question to ask would be: how far from equilibrium would hydrodynamics still be applicable~\citep{Romatschke2017Jan,Akamatsu2021Jan}? Prior studies have explored different ways to address this question using the entropy density~\citep{Campanini2011Sep}, and using the more general question of whether collective flow can persist down to systems of only a dozen final-state particles~\citep{Heinz2019Jul}. In~\citep{Kurkela2019Nov}, the authors tried to quantify the applicability of hydrodynamics in terms of a dimensionless quantity termed \textit{opacity}, which resolved the system sizes into three regimes, viz., \textit{particle-like}, \textit{transition}, and \textit{hydro-like}.

For a long time, hydrodynamics has been associated with a state of thermodynamic equilibrium. As the system under consideration is strictly not in local thermal equilibrium, the term \textit{hydrodynamization} was coined by Casalderrey-Solana et al~\citep{dynamization_coined} to distinguish between the state of a system following low-order viscous hydrodynamic constitutive relations from the state of local thermalization. The usual approach to hydrodynamics involves an expansion of the energy-momentum tensor in terms of gradients of the fields. Here, the fundamental hydrodynamic fields are considered in the local rest frame to build the energy-momentum tensor ($T^{\mu\nu}$) of the system, satisfying the conservation equation ($\partial_\mu T^{\mu\nu} = 0$).
With just the fields in the energy-momentum tensor expression, we have the zeroth-order ideal hydrodynamics. If we include the first-order gradient, which represents the dissipative effects, we run into the causality-violation problem of the Navier--Stokes equation~\citep{acausality}. A few studies have made progress in restoring causality at first order~\citep{Kovtun2019Oct,PhysRevD.104.023015,BDNK2018,BDNK2019}. With the inclusion of the second-order gradient, the formalisms differ on the basis of the number or kind of terms kept in the constitutive relations, resulting in variations, e.g., MIS~\citep{Muller1967Aug_MIS1,Israel1976Sep_MIS2,Israel1979Apr_MIS3}, Denicol--Niemi--Molnar--Rischke (DNMR)~\citep{Denicol2010Apr_DNMR}, aHydro~\citep{ahydro}, and BRSSS~\citep{BRSSS2008}. 

There are signatures that the familiar hydrodynamic gradient expansion, in general, is not a convergent series~\citep{Romatschke2017Jan}. To examine the collective modes of the diverging gradient expansion, we usually consider the linear response of the energy-momentum tensor ($T^{\mu\nu}$) to an arbitrary perturbing source, $S_{\gamma\delta}$~\citep{Florkowski2018Review}, as,
\begin{eqnarray}
 \delta \langle T^{\mu\nu} \rangle (t,{\bf x})  = \int dt \; d^3x \; G^{\mu\nu,\gamma\delta}(t,{\bf x}) \; S_{\gamma\delta},
\end{eqnarray}
where $G^{\mu\nu,\gamma\delta}(t,{\bf x})$ is the two-point retarded correlator, referred to as the Green's function.
The solution of this equation at late times is expressed as, 
\begin{eqnarray}
\delta \langle T^{\mu\nu} \rangle (t,{\bf x})  \sim  e^{-i\omega_{\rm sing}(k) t  + i \textbf{k} \cdot \textbf{x} },
\end{eqnarray}
where $\omega_{\rm sing}(k)$ is the complex singularity of the Fourier-transformed two-point retarded correlator. 
The real part of this complex frequency is termed the ``hydrodynamic mode frequency'' and is responsible for the excitation of the equilibrium plasma. The imaginary part is referred to as the ``non-hydrodynamic mode frequency,'' which causes dissipative effects. For BRSSS theory, these singularities have been worked out for the sound channel~\citep{Spalinski2016} with the hydrodynamic mode frequency given by,
\begin{eqnarray}
\omega^{\pm}_{\mathrm{h}} = \pm \frac{k}{\sqrt{3}} - \frac{2i\eta k^2}{3Ts} + \dots 
\label{eq:hydro_freq}
\end{eqnarray}
and the corresponding non-hydrodynamic mode frequency is given by,
\begin{eqnarray}
\omega_{\mathrm{nh}} = -i \bigg( \frac{1}{\tau_{\pi}} - \frac{4\eta k^2}{3sT} \bigg) + \dots.
\label{eq:non-hydro}
\end{eqnarray}
In the above equations (\ref{eq:hydro_freq}) and (\ref{eq:non-hydro}), $\eta$ is the shear viscosity, $T$ is the temperature, $s$ is the entropy density, $k$ is the wave vector, and $\tau_{\pi}$ is the second-order transport coefficient called the shear relaxation time.
For vanishing wave vector, these frequencies reduce to,
\begin{eqnarray}
\lim_{k\to 0}  \omega^{\pm}_{\mathrm{h}} = 0 \quad ; \quad  \lim_{k\to 0}  \omega_{\mathrm{nh}} = -\frac{i}{\tau_\pi}.
\label{eq:non-hydro_in_limit}
\end{eqnarray}
The fact that $\omega_{\mathrm{nh}}$ cannot be eliminated in the limit of $k\rightarrow 0$ is the defining feature of the non-hydrodynamic mode.

The contemporary understanding of hydrodynamics indicates that the applicability of hydrodynamics is determined exclusively by the relative contribution of the non-hydrodynamic mode to the hydrodynamic mode. The fluid behavior will break down if the contribution of the non-hydrodynamic mode exceeds that of the hydrodynamic mode. The non-hydrodynamic mode is regulated by the second-order transport coefficient called the shear relaxation time, as can be seen from equation~(\ref{eq:non-hydro_in_limit}).

The primary objective of this work is to study the behavior of \textit{elliptic flow}--the second-order momentum anisotropy coefficient, for extreme values of the shear relaxation time in peripheral p-Pb collisions at $\mathrm{\sqrt{s_{NN}} = 5.02}$ TeV. An increase in the deviation of the elliptic flow for extreme values of the shear relaxation time in peripheral collisions would signify a breakdown of hydrodynamic behavior~\citep{Romatschke2017Jan}.
In the previous study~\citep{MyHydroStudy1}, the onset of hydrodynamization for peripheral Pb--Pb collisions at $\mathrm{\sqrt{s_{NN}} = 2.76}$ TeV and Au--Au collisions at $\mathrm{\sqrt{s_{NN}} = 200}$ GeV was found to be roughly at $dN/dy \approx 10$. 

The structure of this paper is as follows. In section~\ref{sec:framework}, a brief description of the models utilized for the different collision stages is given. The choice of parameters and the benchmarking of the model against the available experimental data are explained in section~\ref{sec:parameters}. In section~\ref{sec:results}, the results for the elliptic flow as a function of $p_\text{T}$ and $dN/dy$, for varying values of the shear relaxation time factor, are discussed, followed by a summary and conclusion in section~\ref{sec:conclusion}.

\section{\label{sec:framework} Model framework}
The soft sector of p-Pb collisions at $\mathrm{\sqrt{s_{NN}} = 5.02}$ TeV was simulated using a comprehensive framework called JETSCAPE~\citep{Jetscape2019Framework}. This modular framework has shown promising results in simulating both regimes of dynamics, viz., the soft and hard sectors, as well as their interaction, in recent studies~\citep{JS_jets, JS_Multisystem_Bayesian, JS_suppresion, JS_charm, JS_smallsystems}. The present study utilized JETSCAPE 3.6.6~\citep{JETSCAPEgithub}. The individual stage modules selected for this study are $\mathrm{T_RENTo}$, Freestreaming, MUSIC, iSS, and SMASH. A very brief description of each of these modules is given below:

\subsection{  \label{sec:trento} Initial condition: $\mathrm{T_RENTo}$ }
The $\mathrm{T_RENTo}$ model~\citep{Trento2015} generates the initial energy density deposited in the plane transverse to the motion of the approaching nuclei right after they have crossed each other, without assuming any underlying physical dynamics. To begin with, nucleon positions in the approaching nuclei A and B are randomly sampled from a Woods--Saxon distribution~\citep{WoodsSaxon1954} while ensuring a minimum distance ($d_{min}$) between nucleons to account for repulsive forces between them. The individual nucleon density is assumed to have a Gaussian form,
\begin{eqnarray}
\rho_n (\boldsymbol{x}) = \frac{1}{(2\pi w^2)^{3/2}} \text{exp} \bigg(-\frac{|\boldsymbol{x}|^2}{2w^2}  \bigg),
\label{eq:nucleon_density}
\end{eqnarray}
where the standard deviation, $w$, is a $\mathrm{T_RENTo}$ parameter that signifies the effective nucleon width. The $\mathrm{T_RENTo}$ model treats the constituents inside nucleons in a manner analogous to how nucleons are treated inside a nucleus~\citep{trento_nnGlauber}. Consider a nucleon-nucleon collision at a given impact parameter, $\textbf{b}_n$.
The density of a single nucleon in equation~(\ref{eq:nucleon_density}) is used to find the nucleon thickness function, $ T_{n}(\boldsymbol{x}_{\perp}) = \int dz \; \rho_n (\boldsymbol{x}_{\perp},z)$, which is then used to calculate the nucleon-nucleon overlap function,
\begin{eqnarray}
T_{nn}(\boldsymbol{b}_n) = \int d^2x \; T_{n}(\boldsymbol{x}_{\perp})  T_{n}(\boldsymbol{x}_{\perp} - \boldsymbol{b}_n).
\end{eqnarray} 
The collision probability of at least one nucleon-nucleon pairwise interaction is given by~\citep{Enterria2010Mar},
\begin{eqnarray}
P^{coll}_{nn}(\boldsymbol{b}_n) = 1 - exp [- \sigma^{\text{inel}}_{\text{eff}} T_{nn} (\boldsymbol{b}_n)],
\label{eq:single_coll_prob}
\end{eqnarray}
where $\sigma^{\text{inel}}_{\text{eff}}$ is the effective cross section for interaction between nucleon constituents. Starting with this, we find the \textit{number of participating nucleons} ($N_{\text{part}}$) from both nuclei \footnote{ From the analytical Glauber model, the probability of $n$ \textit{nucleon-nucleon collisions} is calculated using the single nucleon-nucleon collision probability, $P^{coll}_{nn}(\boldsymbol{b}_n)$, from equation~(\ref{eq:single_coll_prob}). Using that, one can find the average number of nucleons that participate in more than one pairwise collision ($N^A_{\mathrm{part}}$ and $N^B_{\mathrm{part}}$) from each nucleus, A and B~\citep{MorelandThesis}.}
The nucleon density given by equation~(\ref{eq:nucleon_density}) is summed over the number of participating nucleons to calculate the nuclear density as,
\begin{eqnarray}
\tilde{\rho}^{\text{part}}_{A} (\boldsymbol{x}) = \sum_{i=1}^{N^A_{\mathrm{part}}} \gamma_i \; \rho_n(\boldsymbol{x} - \boldsymbol{x}_i + \boldsymbol{b}/2)
\label{eq:trento_fluc}
\end{eqnarray}
Here, $N^A_{\mathrm{part}}$ is the number of participating nucleons from nucleus A, $\textbf{\textit{b}}$ is the nucleus-nucleus collision impact parameter, and $\gamma_i$ is a weight factor sampled from a gamma distribution to introduce event-by-event fluctuations similar to those in the experimental setup~\citep{fluctuating_gamma_weight}.
An analogous equation is used for nucleus B with a negative sign for the $\textbf{\textit{b}}/2$ term. This fluctuating nuclear density is then used to calculate the nuclear thickness function, $\tilde{T}_A (\vec{x}_{\perp})  = \int dz \; \tilde{\rho}_A (\boldsymbol{x}_{\perp},z) $. The nuclear overlap function is calculated using a generalized mean called the \textit{reduced thickness},
\begin{eqnarray}
T_R = \bigg( \frac{ \tilde{T}^p_A + \tilde{T}^p_B } {2}  \bigg)^{1/p},
\end{eqnarray}
where $p$ is a dimensionless real-valued parameter. The specific values $p = -1, 0, +1$ correspond to the arithmetic, geometric, and harmonic means, respectively, as stated in \citep{Trento2015}. This \textit{reduced thickness} is then scaled by a normalization factor to obtain the initial energy density deposited after the nucleus-nucleus collision.
\subsection{Pre-equilibrium: freestreaming }
The pre-equilibrium stage is intended to fill the gap between the initial stage of the collision and the near-equilibrium hydrodynamic stage with out-of-equilibrium dynamics. The approaching nuclei cross each other in a time, $\tau \approx  2R / \gamma \beta$, where $R$ is the rest-frame radius of the nucleus, $2R/\gamma$ is the Lorentz-contracted length, and $\beta=v/c$. For the LHC and RHIC, this crossing time is of the order of $10^{-3}$ fm/\textit{c} and $10^{-1}$ fm/\textit{c}, respectively. The crossing time is an appropriate moment to initiate the pre-equilibrium dynamics. However, there is no such rationale for deciding the end time of the pre-equilibrium dynamics or the starting time of hydrodynamics~\footnote{Other than using values that lead to satisfactory agreement between the generated observables and the experimental data, a fiducial value is generally selected over the approximate range 0.2--1.5 fm/\textit{c}.}. The role and effect of the pre-equilibrium stage on the final-stage observables need to be looked at in more detail~\citep{Prehydrodynamic2022}. The Freestreaming module used as a pre-equilibrium stage for this study is based on the collisionless Boltzmann equation of massless particles~\citep{freestream2015ChunShen,JS_Multisystem_Bayesian},
\begin{eqnarray}
p^\mu \partial_\mu f(X;P) = 0  ,
\end{eqnarray}
where $f(X;P)$ is the phase-space momentum distribution of particles in the plane transverse to the beam direction at the time of the collision. This distribution is assumed to be locally isotropic. The solution of the above Boltzmann equation for a boost-invariant system is calculated in terms of a moment of the distribution $f(X;P)$, which is given by,
\begin{eqnarray}
   F[X;\Omega_p] \equiv \frac{g}{(2\pi)^3} \int P_0^3\,  f(X;P) \,  dP_0,
\end{eqnarray}
where $\Omega_p$ is the solid angle in momentum space, $g$ is a degeneracy factor, and $P^0 = |\boldsymbol{P}|$. We are interested in calculating the energy-momentum tensor for this kinetic theory, which will be evolved for the Freestreaming duration and then matched with the fluid energy-momentum tensor at the start of hydrodynamics. The Freestreaming energy-momentum tensor is expressed using the above moment as,
\begin{eqnarray}
\label{Tmn_fs}
    T^{\mu\nu}(t,\boldsymbol{x}) = \int  \hat{p}^{\mu} \hat{p}^{\nu}\, F\bigl[ t_0, \boldsymbol{x}{-}\boldsymbol{v}(t{-}t_0 ) ; \Omega_p \bigl]\,  d\Omega_p,
\end{eqnarray}
where, $t_0$ is the time at which we start the Freestreaming module.

The initial condition for $T^{\mu\nu}$ is set through the initial energy density obtained from the $\mathrm{T_RENTo}$ module as,
\begin{eqnarray}
   T^{00}(t_0, \boldsymbol{x}) = \epsilon =  \mathcal{N} F(t_0, \boldsymbol{x}),
\end{eqnarray}
where $\mathcal{N}$ is $2\pi$ for the 2-dimensional momentum distribution. The evolution is stopped at the Freestreaming time, $\tau_{\mathrm{fs}}$, which is also the time when the hydrodynamic module is initiated. The complete $T^{\mu\nu}$ of the pre-equilibrium stage is smoothly and consistently matched to the fluid stage, in the local rest frame, using the Landau matching condition,
\begin{eqnarray}
 u_{\mu}T^{\mu}_{\nu} = \epsilon u_{\nu}.
 \label{eq:landau_matching}
\end{eqnarray}
This includes matching the viscous dissipative part of the tensor as well. The Freestreaming module is equipped with a feature to use a centrality-dependent Freestreaming time, $\tau_{\mathrm{fs}}$, given by,
\begin{eqnarray}
\tau_{\mathrm{fs}} = \tau_R \bigg( \frac{\langle \bar{\epsilon}_{\text{cent}} \rangle }{\bar{\epsilon}_R} \bigg)^\alpha.
\end{eqnarray}
Here, $\langle \bar{\epsilon}_{\text{cent}} \rangle$ is the average transverse energy density of a given centrality, $\tau_R$ is the normalization used for the duration for which Freestreaming operates, $\alpha$ controls the contribution of the initial energy density, and $\bar{\epsilon}_R$ is an arbitrary energy density reference. 
\subsection{Viscous hydrodynamics: MUSIC }
MUSIC~\citep{music2010} is a second-order dissipative fluid-dynamics code based on the DNMR theory~\citep{Denicol2012Jun_MUSIC1,Denicol2015Feb_MUSIC2}, which resums the comprehensive second-order BRSSS equations~\citep{BRSSS2008}. MUSIC uses the Landau choice of the velocity rest frame, the same as in equation~(\ref{eq:landau_matching}). It has been widely used in phenomenological studies for various collision systems and energies~\citep{2020gamut,ShenYan2020}.
MUSIC shares the non-hydrodynamic mode of the BRSSS formalism, with the non-hydrodynamic mode frequency given in equation~(\ref{eq:non-hydro_in_limit}), as stated in section~8.3 of~\citep{Denicol2012Jun_MUSIC1}. 
An equation of state that interpolates between the HotQCD~\citep{lqcdEoS1} lattice predictions at high temperatures and the Hadron Resonance Gas model predictions at lower temperatures is used~\citep{HuovinenPetreczky2010}.
The energy-momentum tensor conservation equations, $\partial_{\mu} T^{\mu\nu} = 0$, in MUSIC are solved using the Kurganov-Tadmor algorithm~\citep{music2010,KurganovTadmor2000}. The fluid energy-momentum tensor is constructed using the ideal fluid part, $T^{\mu\nu} = \epsilon u^\mu u^\nu + \Delta^{\mu\nu} P$, and a dissipative part, $\Pi^{\mu\nu}$. It is customary to further split this dissipative tensor into traceless and trace parts, $\Pi^{\mu\nu} = \pi^{\mu\nu} + \Pi \Delta^{\mu\nu}$, which are referred to as the shear and bulk parts of the dissipative tensor, respectively. The term, $\Delta^{\mu\nu} \equiv g^{\mu\nu} - u^{\mu}u^{\nu}$, projects onto the spatial part in the local rest frame. For MUSIC, the evolution equation for the shear tensor is given by:
\begin{align}
    \tau_{\pi }\dot{\pi}^{\left\langle \mu \nu \right\rangle }+\pi ^{\mu \nu }
    &= 2\eta \sigma ^{\mu \nu }-\delta _{\pi \pi }\pi ^{\mu \nu }  \partial_\lambda u^\lambda
    +\varphi_{7}\pi _{\alpha }^{\left\langle \mu \right. }\pi ^{\left. \nu \right\rangle \alpha } \notag 
    \\
    &\ -\tau _{\pi \pi }\pi _{\alpha }^{\left\langle \mu \right. }\sigma^{\left. \nu \right\rangle \alpha }+\lambda _{\pi \Pi }\Pi \sigma ^{\mu\nu}.
\label{relax_eqn_pi}
\end{align}
The evolution equation for the bulk term is given by,
\begin{align}
    \tau_{\Pi }  \dot{\Pi}  +\Pi &= -\zeta \partial_\lambda u^\lambda -\delta _{\Pi \Pi }\Pi  \partial_\lambda u^\lambda
    + \lambda _{\Pi \pi }\pi ^{\mu \nu }\sigma_{\mu \nu }\;,
\label{relax_eqn_PI}
\end{align}
where $\dot{\Pi} = u^\lambda \partial_\lambda \Pi$, $\dot{\pi}^{\langle\mu\nu\rangle} = \Delta^{\mu\nu}_{\alpha\beta} u^\lambda \partial_\lambda \pi^{\alpha\beta}$, and $\sigma^{\mu\nu} = \Delta^{\mu\nu}_{\alpha\beta}\partial^\alpha u^\beta$, where $\Delta_{\alpha\beta}^{\mu\nu}$ is a projector that extracts the part of a rank-2 tensor that is traceless, symmetric, and orthogonal to the velocity $u^\mu$, and is given by:
\begin{eqnarray}
\Delta^{\mu\nu}_{\alpha\beta} \equiv \frac{1}{2}(\Delta^{\mu}_{\alpha} \Delta^{\nu}_{\beta} + \Delta^{\nu}_{\alpha} \Delta^{\mu}_{\beta}) - \frac{1}{3}\Delta^{\mu\nu}\Delta_{\alpha\beta}.
\label{eq:projector2}
\end{eqnarray}
Here, $\delta _{\Pi \Pi }$, $\lambda _{\Pi \pi }$, $\delta _{\pi \pi }$, $\varphi _{7}$, $\tau _{\pi \pi }$, $\lambda_{\pi\Pi }$, $\tau _{\pi }$, and $\tau _{\Pi }$ are the second-order transport coefficients, which depend on the details of the underlying microscopic theory under consideration~\citep{Denicol2014Nov_review}.

The first-order transport coefficients, shear and bulk viscosities, are conventionally introduced by taking their ratio with the entropy density. In a realistic scenario, the shear viscosity to entropy density ratio ($\eta/s$) would be a function of temperature. Constraints on this temperature-dependent $\eta/s$ have been explored in a Bayesian study~\citep{eta_by_s_NATUREpaper}, motivated in part by the conjectured lower bound on $\eta/s$ from gauge-gravity duality~\citep{KSS_limit}. The temperature-dependent $\eta/s$ in this work has been parameterized as~\citep{JS_Multisystem_Bayesian}, 
\begin{eqnarray}
    \frac{\eta}{s}(T) &=& a_{\rm low}\, (T-T_{\eta})\, \Theta(T_{\eta}-T)+ (\eta/s)_{\rm kink} \nonumber\\
    && +\, a_{\rm high}\, (T-T_{\eta})\, \Theta(T-T_{\eta}).
\label{eq:eta_by_s_T}
\end{eqnarray}
Here, $\Theta$ is the Heaviside step function, and $a_{\rm high}$, $a_{\rm low}$, $(\eta/s)_{\rm kink}$, and $T_{\eta}$ are parameters chosen such that $\eta/s$ stays within the bounds allowed by the Bayesian analysis. The values of these parameters are given in section~\ref{sec:parameters}. The second-order transport coefficient called the \textit{shear relaxation time}, $\tau_\pi$, is related to $\eta/s$ as,
\begin{equation}
    \tau_\pi(T)= \frac{b_{\pi}}{T} \frac{\eta}{s}(T)
\label{eq:relax_time}    
\end{equation}
where $b_{\pi}$ is a constant whose value has been estimated from microscopic theories. We will refer to this coefficient, $b_{\pi}$, as the \textit{shear relaxation time factor} in this study. 
\subsection{Hadron afterburner: (iSS + SMASH) }
In actual collisions, when the system cools, the QGP will cease to exist below the pseudocritical temperature of 156 MeV (from a lattice QCD estimate~\citep{Ding2021Jan,BazavovChiralCrossover2019}).
The quarks will then hadronize at around the chemical freezeout temperature.
In the modeling, for simplicity, the hadronization process is bypassed, and the direct formation of hadrons is considered at the end of the hydrodynamic stage using the Cooper--Frye prescription.
Since the degrees of freedom are changing from hydrodynamic fields to hadrons, this transition point is referred to as \textit{particlization}.  
When the fluid cell temperature drops below the particlization temperature, a hypersurface of fluid cells is created. A hypersurface is a spacetime fluid volume  that has achieved the particlization temperature in its evolution. The hydrodynamic fields at this hypersurface are converted into particles using the Cooper--Frye prescription~\citep{Cooper_frye},
\begin{eqnarray}
E\frac{d^3N_i}{dp^3} = \frac{dN_i}{p_\text{T} dp_\mathrm{T} dy d\phi_p} = g_i \int_\Sigma f_i( u^\mu p_\mu ) p_\mu  d{\Sigma_\mu}.
\label{eq:cooper_frye}
 \end{eqnarray}
The left-hand side of this equation provides the momentum spectra of particles of species $i$ at the hypersurface denoted by $\Sigma$, in the direction normal to the hypersurface. The coefficient $g_i$ is the degeneracy factor of species $i$.
The quantity $f_i( u^\mu p_\mu )$ is the phase-space distribution of the particles generated. 
The information about the hydrodynamic fields is contained in the flow velocity $u^\mu$ and in the differential surface-element vector, $d\Sigma_\mu$.
The hadrons produced are in a near-equilibrium state, and hence the distribution functions going into equation~(\ref{eq:cooper_frye}) are chosen to be Fermi--Dirac and Bose--Einstein,
\begin{eqnarray}
f( u^\mu p_\mu ) = \frac{1}{(2\pi)^3} \frac{1}{ \text{exp}\big[(  u^\mu p_\mu  - \mu_i ) / T_{\mathrm{cfo}} \big] \pm 1},
\label{eq:local_equilibrium}
\end{eqnarray}
where $\mu_i$ is the equilibrium chemical potential of species $i$, and $T_{\mathrm{cfo}}$ is the chemical freezeout temperature or particlization temperature. If other charges are considered, their chemical potentials can be obtained by summing the products of their corresponding charges and chemical potentials. To capture the out-of-equilibrium dissipative nature of the fluid as translated into the particle description, a correction to this phase-space distribution should be added.
 
In the framework of this study, particles are sampled from random positions on the hypersurface, which is handled by the iSS package~\citep{iSSpaper}.
To save computational time, particles are usually sampled more than once from the hypersurface, which also facilitates the estimation of observables with smaller fluctuations. Since the multiplicity is lower in peripheral collisions as compared to central ones, more ``oversampling'' is required for peripheral collision events in order to cover the hypersurface. 
The estimated observables have to be appropriately normalized by the number of oversampling events.
Another package called SMASH~\citep{SMASH} handles particle scattering in the expanding fireball and resonance decays. For this, it solves a tower of Boltzmann equations,
\begin{eqnarray}
p^\mu \partial_\mu f( u^\mu p_\mu )  = C[f_i].
\end{eqnarray}
Here, $ C[f_i]$ is the collision term that encodes the scattering of particles of species $i$. All soft sector observables can be calculated from the energy and momenta of particles produced from the afterburner stage.

\section{\label{sec:parameters}Benchmarking and setting model parameters}

The fluid description stops at particlization, and it turns out that the dissipative correction to the local-equilibrium distribution for particles in equation~(\ref{eq:local_equilibrium}) is non-trivial. This correction is a major source of uncertainty. A recent Bayesian study~\citep{JS_Multisystem_Bayesian} for nucleus-nucleus collisions compared three choices for this correction, namely, Grad's 14-moment method~\citep{Grad1949Dec}, the first-order Chapman-Enskog expansion in the relaxation-time approximation~\citep{Chapman1939,Chapman_Enskog}, and the Pratt-Torrieri-Bernhard's modified equilibrium distribution~\citep{Pratt_Torrieri,Bernhard2018Apr}. It is concluded that the light-particle observables favor Grad's and Pratt-Torrieri-Bernhard's correction terms. For the present study, we chose the set of constrained parameters centered around Grad's method correction term, as given in table~2 of~\citep{JS_Multisystem_Bayesian}. Hence, we are assuming that the constrained parameter set for Pb--Pb can also be approximately utilized for the p-Pb system.
The MUSIC parameters related to the $\eta/s$ parameterization in equation~(\ref{eq:eta_by_s_T}) are set to: $a_{\rm high}=$ 0.37 GeV, $a_{\rm low}=$ -0.776 GeV, $(\eta/s)_{\rm kink}=$ 0.096 GeV, and $T_{\eta}=$ 0.223 GeV. A parametrization similar to equation~(\ref{eq:eta_by_s_T}) for the bulk viscosity to entropy density ratio ($\zeta/s$) and the corresponding parameters can be found in~\citep{JS_Multisystem_Bayesian}. As the p-Pb collisions have a shorter lifetime, we chose a slightly lower value of the Freestreaming time, $\tau_{\mathrm{fs}}=$ 1 fm/\textit{c}. Also, a realistic particlization temperature of 150 MeV has been used, which is the chemical freezeout temperature~\citep{chem_freezeout_temp}. 
The input files for the framework and individual models have been made available at~\citep{github_inputs}.
The observables are produced for the kinematic cuts: $0<p_\text{T}<3$ GeV/\textit{c} and $-0.8<y<0.8$. The nucleon-nucleon collision cross section has been set to $\sigma_{NN}$ = 67 mb~\citep{Improved_MonteCarlo}. The shear relaxation time factor ($b_\pi$) of equation~(\ref{eq:relax_time}) is set to 2 and 6 for the elliptic flow results of this study, which is a bound well within the causality regime~\citep{Pu2010Jun,JS_Multisystem_Bayesian}.

\begin{figure}
	\includegraphics[scale=0.39]{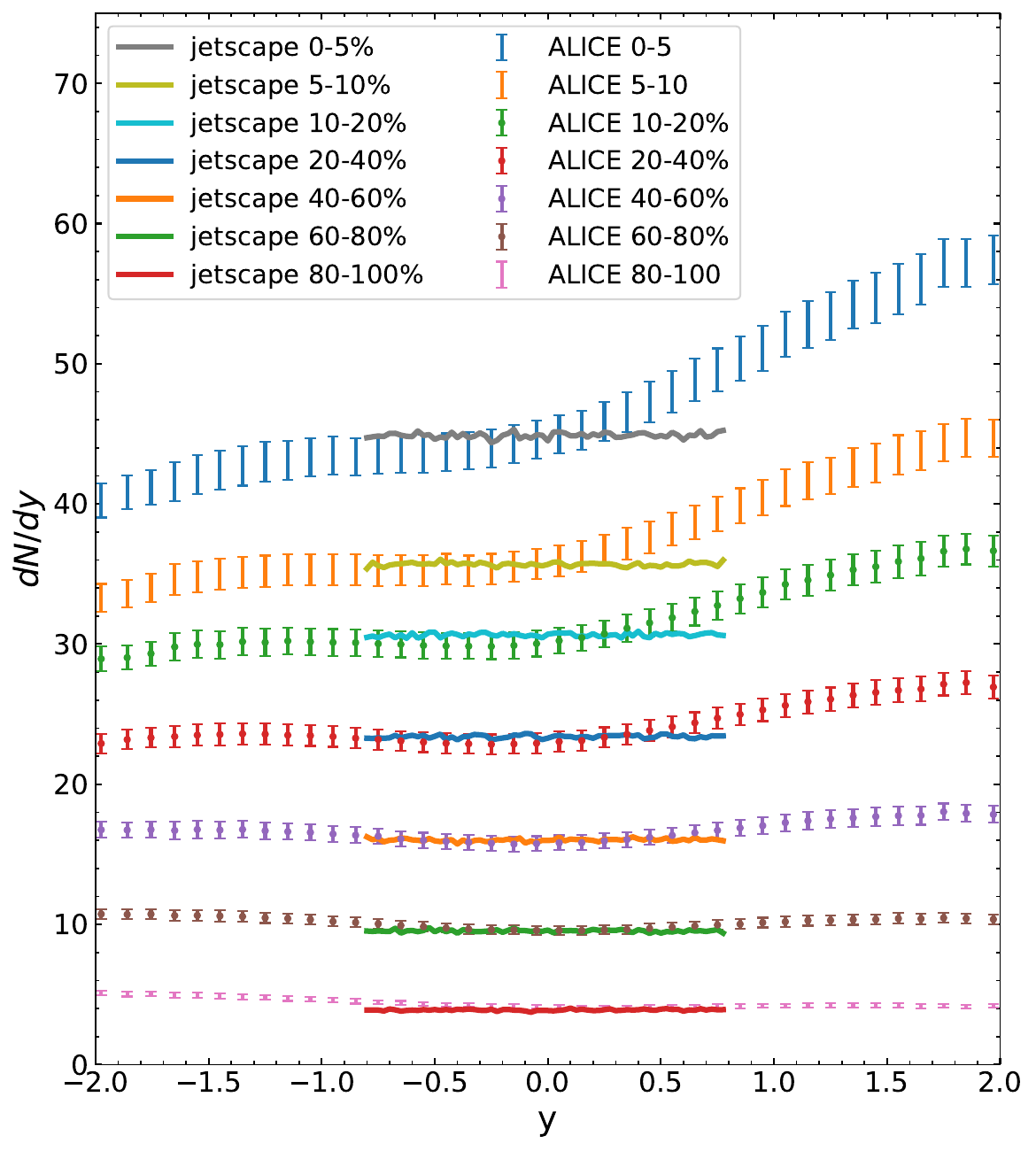}
	\caption{Rapidity spectra of charged particles ($\pi^{\pm}$, $K^{\pm}$, $p$) for the mentioned centrality classes. The experimental pseudorapidity spectra values, taken from \citep{expt_rapSpec}, are available over a wider rapidity range, but the simulated results for the 2+1D system are only valid at midrapidity.}
	\label{fig:rap_spectra}
\end{figure}

The near-central collision rapidity spectra are asymmetric for p-Pb collisions, and ideally, the full 3+1D hydrodynamics should be used to model the dynamics of such a system.
However, the elliptic flow, rapidity, and $p_\text{T}$ spectra obtained from 2+1D hydrodynamics act as decent approximations for those obtained from 3+1D hydrodynamics~\citep{Shen2017Jan}. 
The initial-state energy-density normalization is varied to match the generated charged-particle rapidity spectra with the experimental data for seven centrality classes, as shown in figure~\ref{fig:rap_spectra}. The charged particles considered for analysis are pions, kaons, and protons. Figure~\ref{fig:rap_spectra} also shows a plateau at the midrapidity region for peripheral collisions, indicating that the expansion dynamics are boost invariant~\citep{Bjorken1983Jan}. As the system is 2+1D boost invariant, the obtained spectra are compared with the experimental data near midrapidity. A good agreement is noticeable for the peripheral collisions, which is the regime of interest for this study.
   
\begin{figure}
	\includegraphics[scale=0.4]{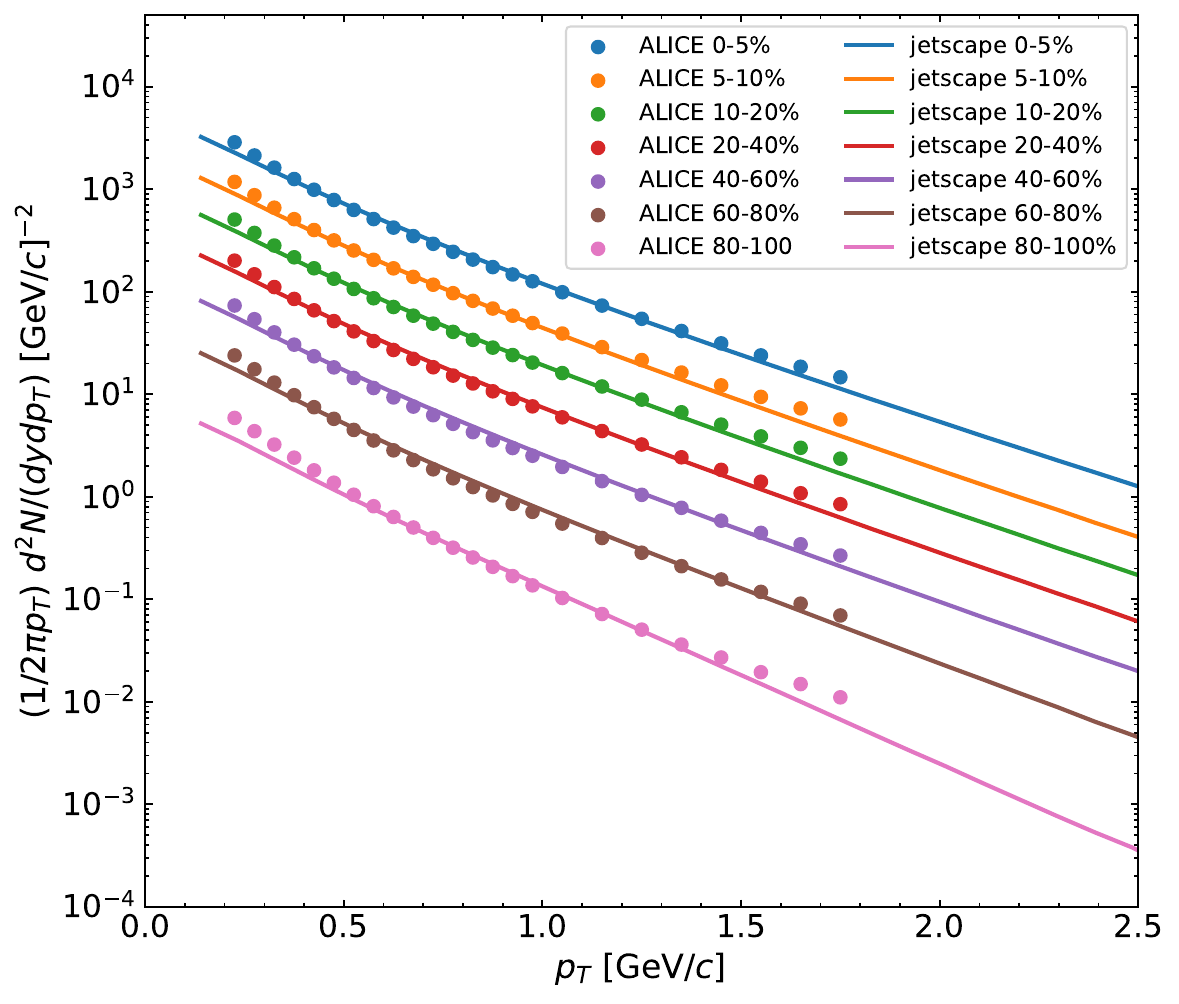}
	\caption{Transverse momentum spectra of charged pions ($\pi^+ + \pi^-$) for the mentioned centrality classes. Experimental data are taken from \citep{Abelev2014Jan}.}
	\label{fig:pT_spectra}
\end{figure}

Figure~\ref{fig:pT_spectra} compares the experimental $p_\text{T}$ spectra for charged pions in the range of $0.15 < p_\text{T} < 1.8$ GeV/\textit{c} with the corresponding simulated data produced in the range of $0.15 < p_\text{T} < 2.5$ GeV/\textit{c}.
The agreement with the experimental data is satisfactory within the $p_\text{T}$ range shown. 
The agreement degrades at low $p_\text{T}$ because the softest particles are most sensitive to freeze-out/radial-flow and resonance-decay systematics.
The simulated data begins to deviate from the experimental spectra beyond $p_\text{T} > 1.5$ GeV/\textit{c} due to viscous corrections ($\delta f$) in the phase-space distribution of equation~(\ref{eq:local_equilibrium})~\citep{Teaney_2002aj}.

\begin{figure}
	\includegraphics[scale=0.35]{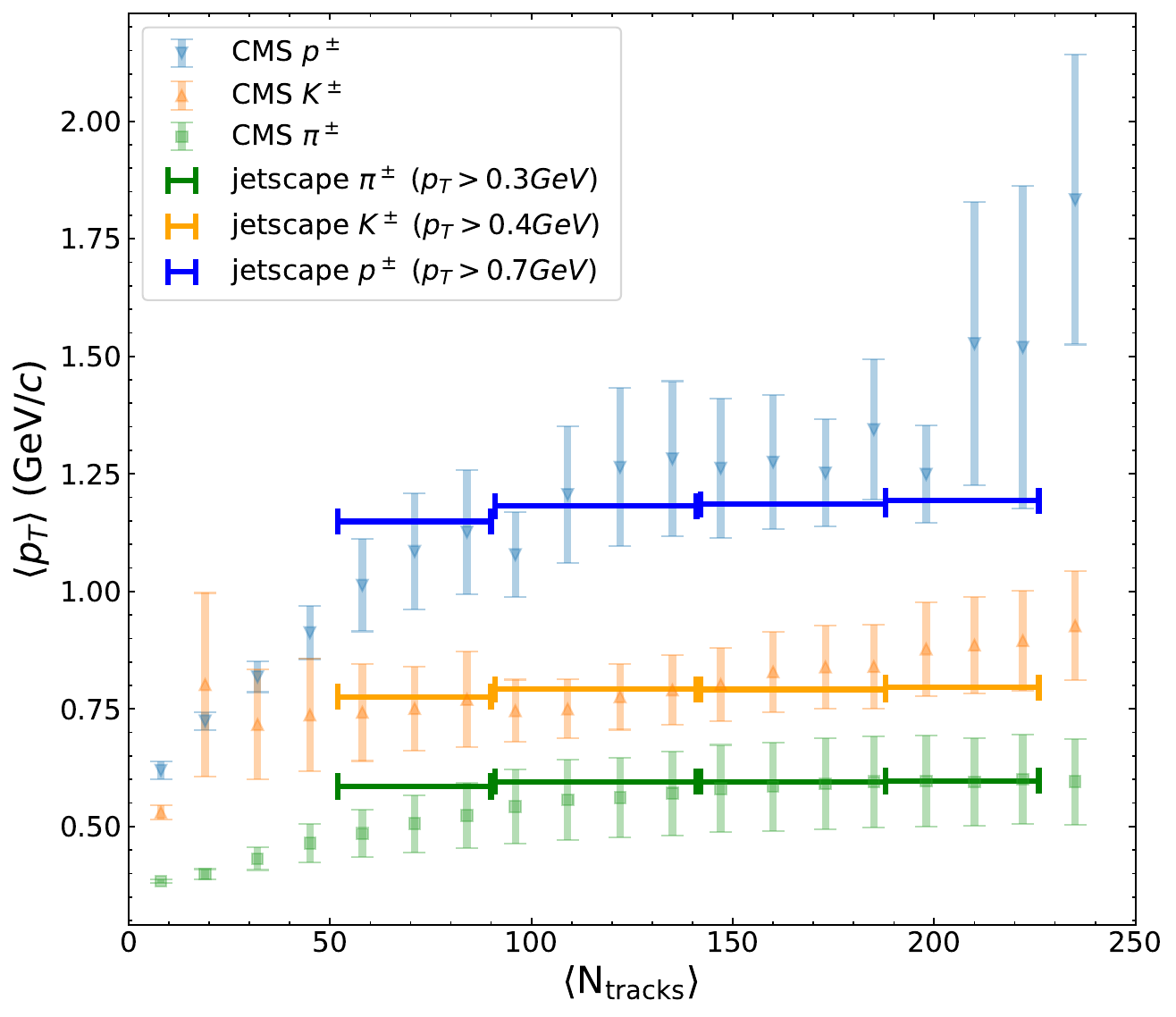}
	\caption{Mean transverse momentum as a function of the average number of corrected tracks, $\langle N_{\text{tracks}} \rangle$. Experimental data are taken from \citep{expt_avgpT}.}
	\label{fig:avg_pT}
\end{figure}

Figure~\ref{fig:avg_pT} shows the comparison of the mean $p_\text{T}$ with the corresponding experimental data as a function of centrality.
The obtained mean $p_\text{T}$ roughly agrees with the experimental data for lower cuts on $p_\text{T}$ of 0.3, 0.4, and 0.7 GeV/\textit{c} for pions, kaons, and protons, respectively.   

The simulated mean $p_\text{T}$ agrees with the data to within about 10\%--15\% for pions and kaons over the full range of $\langle N_{\mathrm{tracks}} \rangle$ studied.
For protons, the agreement is comparable in the more peripheral bins, but since the simulated mean $p_\text{T}$ is nearly independent of centrality over the range studied here while the data continue to rise with multiplicity, the simulation increasingly underestimates the proton data toward more central collisions, with the deviation growing to roughly 25\%--30\% in the most central bins.

The centrality values are set through percentile centrality bins in $\mathrm{T_RENTo}$, following the centrality-class conventions established for p-Pb collisions at this energy~\citep{ALICECollaboration2015Jun}. The centrality of the experimental data is available in terms of the average number of corrected tracks, $\langle N_{\mathrm{tracks}} \rangle$. The correlation between $\langle N_{\mathrm{tracks}} \rangle$ and centrality percentiles is obtained from table~1 of~\citep{Abelev2014Jan}.

The experimental data points for the rapidity and transverse momentum spectra as well as for mean-$p_\text{T}$ in figures~\ref{fig:rap_spectra}--\ref{fig:avg_pT} respectively, include both statistical and systematic uncertainties, added in quadrature, as reported in the corresponding experimental publications~\citep{expt_rapSpec,Abelev2014Jan,expt_avgpT}; these are typically at the few-percent level for pions and kaons and grow to roughly 10\% or more for protons and for the more peripheral centrality classes, where the available statistics are more limited.

Our simulation results do not include a propagated estimate of systematic uncertainty, as the data points are calculated at exact values of dependent variables ($p_T$, $y$ and centrality class). We have also neglected the statistical errors in the simulated data, as these are significantly smaller compared to the experimental counterparts.
It is nonetheless possible to introduce numerical uncertainty through the choice of the pre-equilibrium (Freestreaming) duration, the particlization prescription, and the Bayesian-constrained parametrizations of $\eta/s(T)$ and $\zeta/s(T)$ (see section~\ref{sec:conclusion}), although at a cost of compute time.

\begin{figure*}
\includegraphics[scale=0.41]{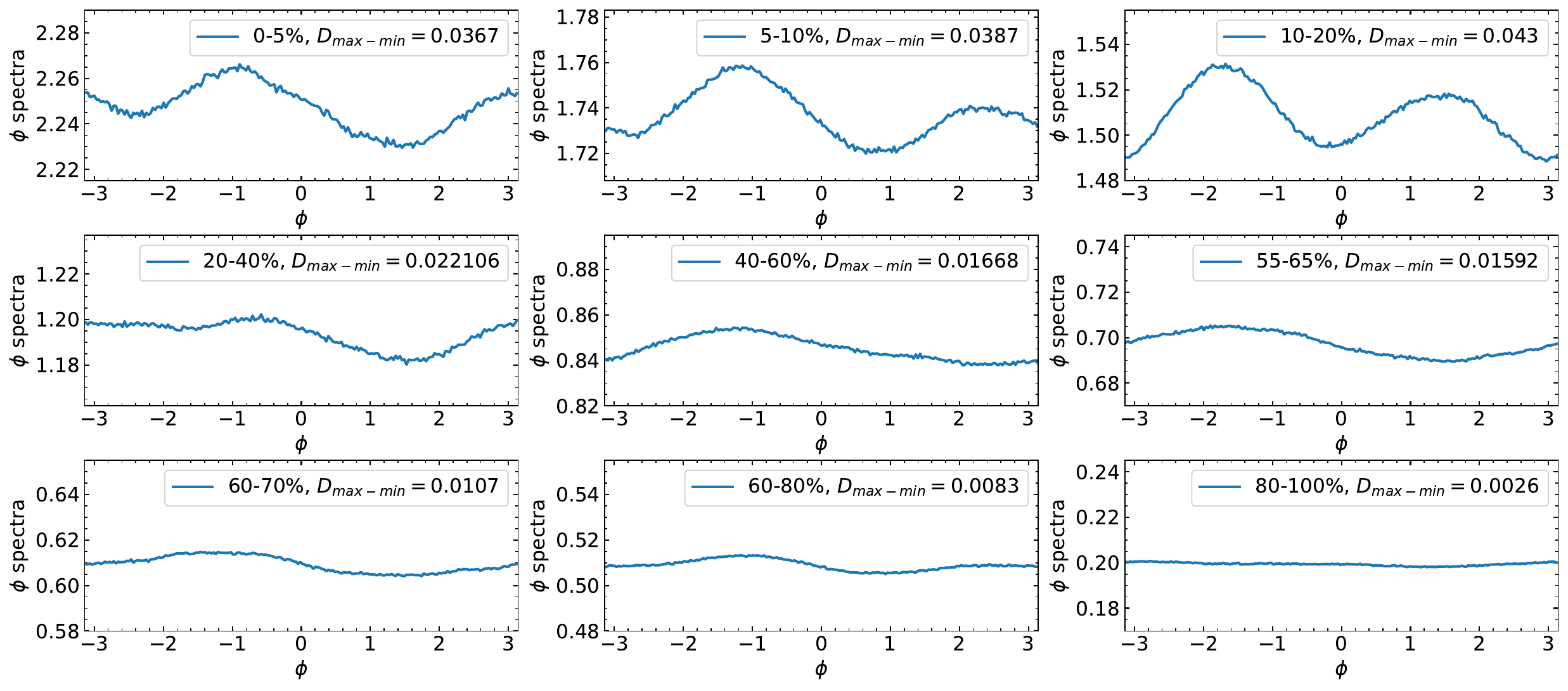}
\caption{Charged-particle azimuthal-angle ($\phi$) distribution for nine centrality classes.}
\label{fig:phi_histo}
\end{figure*}

The number of events, \textit{oversampling}\footnote{Oversampling is a computational optimization strategy available in JETSCAPE, where a large number of events can be generated from a single hypersurface (after the freezeout stage) at each hydrodynamic event.} events, and initial energy density normalization for the nine centrality classes are given in table~\ref{tab:1}.
The number of events selected is the minimum number required to ensure that the generated $\phi$ distributions in figure~\ref{fig:phi_histo} are smooth and would not change significantly with further increase in the number of events. 
The oversampling events are used to minimize the fluctuations in the $\phi$ distributions.
For the mentioned number of oversampling events, we found that the size of the fluctuation in the $\phi$ distribution is significantly smaller than the amplitude of its sine-like curve.

Figure~\ref{fig:phi_histo} shows the $\phi$ distribution of the considered particles for different centrality classes. The separation between the minima and maxima of the curve has also been shown, labeled as $D_{max-min}$. This separation is a measure of the minimum\footnote{The quantity  $D_{max-min}$ corresponds to the ``minimum'' spatial anisotropy because the $\phi$-spectra for each centrality bin are summed over many oversampling events and hence represent a conservative lower bound.} spatial anisotropy for a given centrality bin.
This spatial anisotropy is small for most central collisions, where, ideally, there is azimuthal isotropy even for an asymmetric collision system like p-Pb.
The spatial anisotropy increases noticeably for the centrality class of 10\%--20\%, which is the maximum among all classes. This centrality class also corresponds to the largest magnitude of elliptic flow, as can be seen in figure~\ref{fig:v2_pT_2relax}.
 Beyond this point, it decreases gradually, which is an expected behavior.


\begin{table*}[t]
	\centering
	\begin{tabular}{|>{\bfseries}l|c|c|c|c|c|c|c|c|c|}
		\hline 
		centrality class  & \;  \textbf{0-5\%}  & \; \textbf{5-10\%}  &  \; \textbf{10-20\%}   &  \;  \textbf{20-40\%}   &  \;  \textbf{40-60\%} & \;  \textbf{55-65\%}  &  \; \textbf{60-70\%}  &  \textbf{60-80\%}  \; &  \textbf{80-100\%}\\
		\hline
		$\mathrm{T_RENTo}$ norm  & 17.2  & 17.5   &   20   &   25   &   30.5   &  35.7  &   38.35  &  41   &   82    \\
		\hline
		number of events & 1800   &   1800     &   1800    &  1800     &   1800     &  2600   &  3000   &  3000   & 3000   \\
		\hline
		oversampling  & 1400   &   1400     &   1400    &  1400     &   1600     &  2000   &  2000   &  2000   & 2200   \\
		\hline
	\end{tabular}
	\caption{Centrality percentile classes, the normalization used in $\mathrm{T_RENTo}$, and the number of events and oversampling events for each centrality bin are provided. Events with centrality near 100\% would break the hydrodynamic simulation; hence, for practical purposes, we have selected 85\%--95\% as the representative bin for the 80\%--100\% centrality class.}
	\label{tab:1}
\end{table*}


\section{\label{sec:results}Results and discussion}
Long-range azimuthal anisotropies and elliptic-flow harmonics have been measured experimentally in p-Pb collisions at $\sqrt{s_{NN}} = 5.02$ TeV~\citep{expt_pT_v2}, and the centrality dependence of the bulk transverse-energy production in the same system has likewise been characterized~\citep{dNdy_Npart_expt}; together, these measurements provide the experimental motivation for examining the onset of elliptic flow as a function of both $p_\text{T}$ and rapidity density in the present simulation.
\begin{figure*}
\includegraphics[scale=0.55]{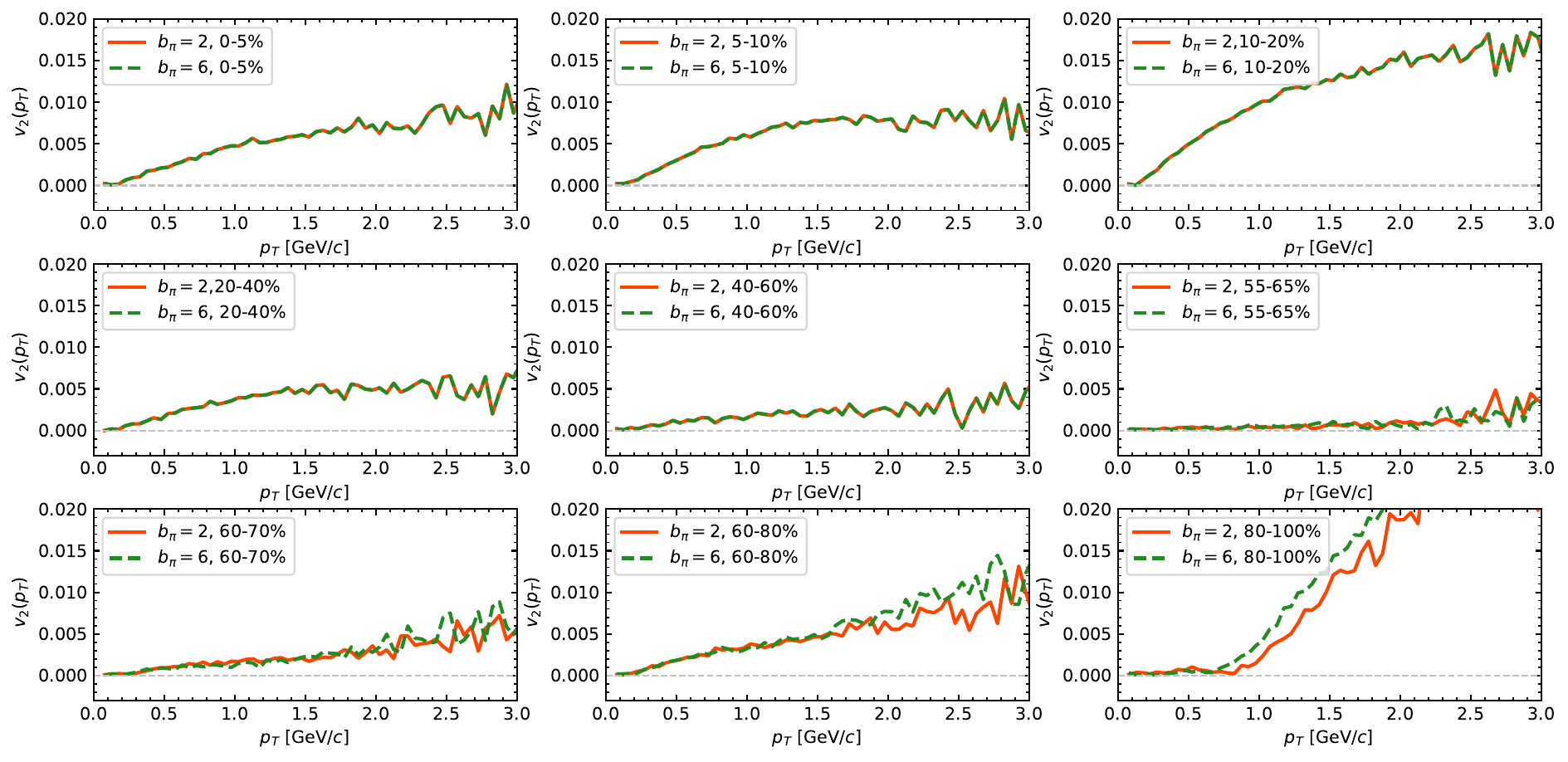}
\caption{Elliptic flow as a function of transverse momentum, for two extreme values of the shear relaxation time factor, over a wide range of centrality classes.}
\label{fig:v2_pT_2relax}
\end{figure*}

For the onset analysis, we generate the charged-particle elliptic flow $v_2(p_\mathrm{T})$ as~\citep{Romatschke2015Jul,PoskanzerVoloshin1998}
\begin{eqnarray}
v_2(p_\mathrm{T}) = \sqrt{s_2(p_\mathrm{T})^2 + c_2(p_\mathrm{T})^2}
\label{eq:elliptic_1}
\end{eqnarray}
where,
\begin{eqnarray}
s_2(p_\mathrm{T}) = \frac{\sum^\text{ch. parti.}_{\text{in } p_\mathrm{T} \text{ bin}} \sin(2\phi) } {\sum^\text{ch. parti.}_{\text{in } p_\mathrm{T} \text{ bin}}}    \nonumber\\
\text{and}  \quad c_2(p_\mathrm{T}) = \frac{\sum^\text{ch. parti.}_{\text{in } p_\mathrm{T} \text{ bin}} \cos(2\phi) } {\sum^\text{ch. parti.}_{\text{in } p_\mathrm{T} \text{ bin}}}.
  \label{eq:elliptic_2}
\end{eqnarray}
Here, $\phi = \arctan(p_y/p_x)$ is the azimuthal angle. The summation ($\sum$) in equation~(\ref{eq:elliptic_2}) is carried over a rapidity window of $-0.8<y<0.8$ and at a specific $p_\text{T}$ bin for all events. In the denominator of the same equation, this summation is just the number of charged particles in that particular $p_\text{T}$ bin.

Figure~\ref{fig:v2_pT_2relax} shows the elliptic flow for nine centrality classes, for the upper and lower bounds on the shear relaxation time factor.
We observe that the flow curves for the two shear relaxation time factors coincide up to the centrality class of 40\%--60\%, which means that the dynamics is insensitive to changes in the shear relaxation time. 
For the centrality class of 55\%--65\%, we notice that the flow curves deviate slightly, and the deviation increases further toward peripheral collisions, up to the most peripheral centrality class of 80\%--100\%.
We also notice that the point of mismatch moves toward lower $p_\text{T}$ from mid-central to peripheral collisions.
This observation is in line with those in the analysis performed for Pb--Pb and Au--Au collisions in~\citep{MyHydroStudy1}.
The magnitude of the elliptic flow is close to zero for the centrality class of 55\%--65\% and increases anomalously for more peripheral centrality classes.

\begin{figure}
	\includegraphics[scale=0.45]{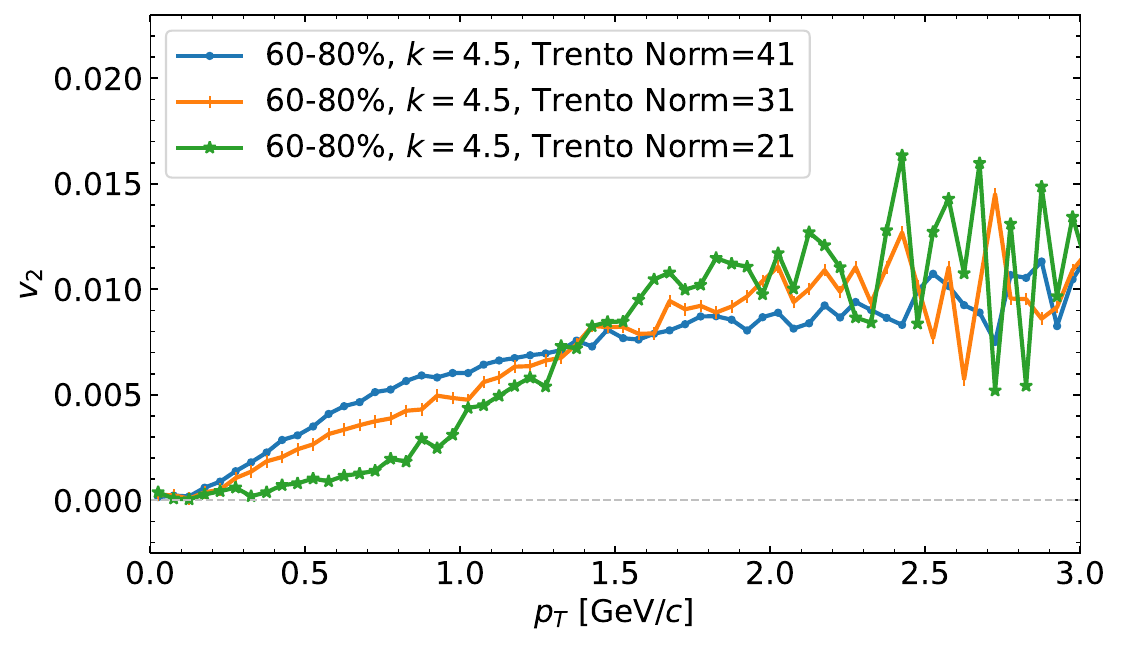}
	\caption{Charged-particle elliptic flow as a function of transverse momentum for the 60\%--80\% centrality bin, with varying initial energy-density normalization while keeping a smooth initial state ($k=4.5$).}
	\label{fig:v2_pT_60_80_varying_norm}
\end{figure}

The peripheral collisions, especially for small systems like p-Pb, are dominated by low particle multiplicity, which heightens the statistical noise and overwrites the collective elliptic flow signal, if any.
In order to probe whether the fluctuations in the produced elliptic flow for peripheral collisions are due to a fluctuating initial state or due to a dilute system, in figure~\ref{fig:v2_pT_60_80_varying_norm} we plot the elliptic flow for the 60\%--80\% centrality class with a smooth initial state but varying $\mathrm{T_RENTo}$ normalization---which controls the magnitude of initial energy density deposited.
The initial-state spatial fluctuations are controlled by the parameter $k$, associated with the $\gamma_i$ term in equation~(\ref{eq:trento_fluc}). Large values of $k$ correspond to a smooth initial state.
We notice in figure~\ref{fig:v2_pT_60_80_varying_norm} that, for decreasing energy-density normalization value, the fluctuations in $v_2$ increase in magnitude, especially at higher $p_\text{T}$, signifying that these $v_2$ fluctuations are not caused by fluctuations in the initial state.

\begin{figure}
	\includegraphics[scale=0.3]{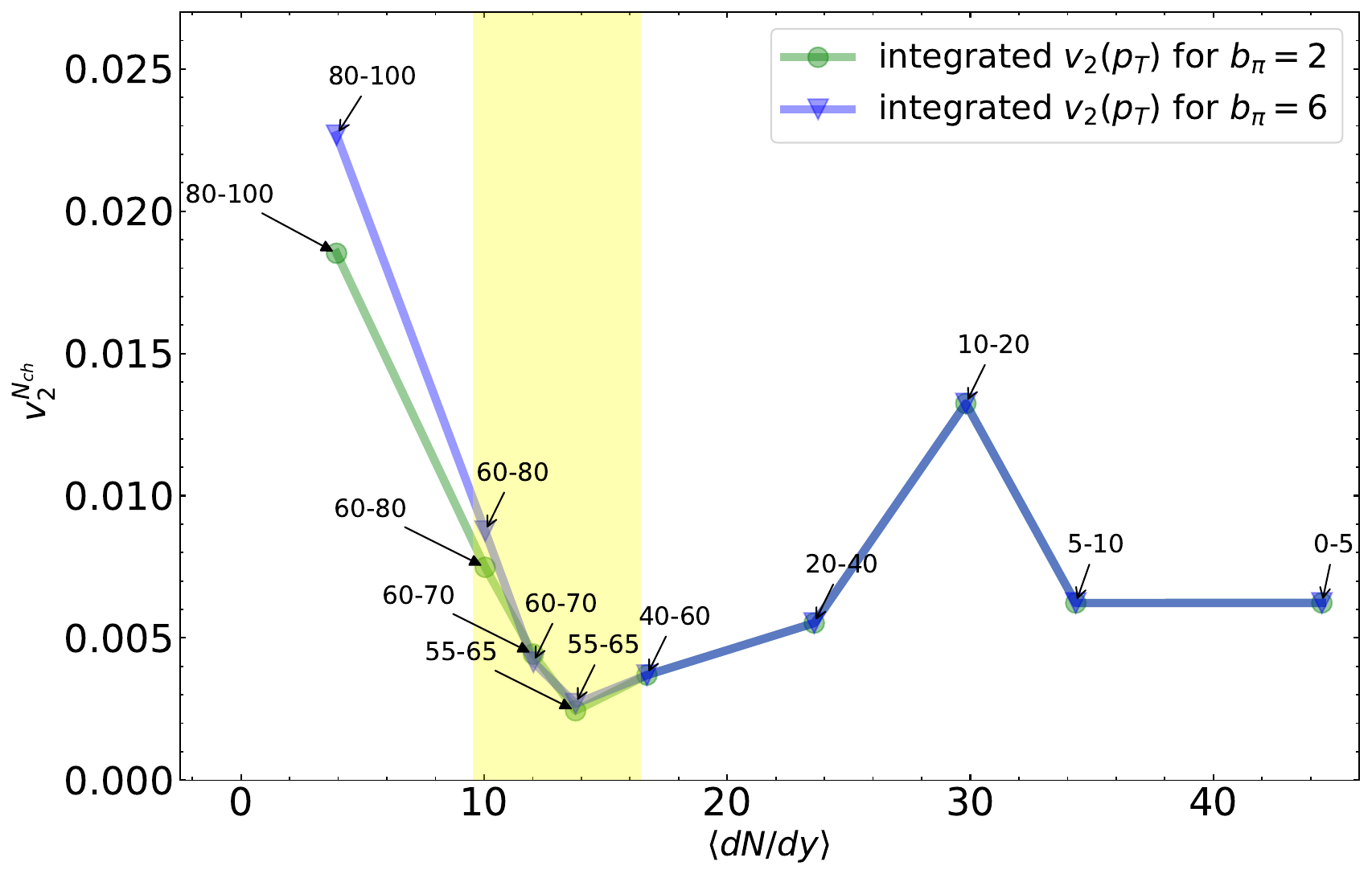}
	\caption{Elliptic flow as a function of rapidity density for the mentioned values of the shear relaxation time factor. The data points are labeled with their centrality class values. The yellow patch represents the approximate onset of hydrodynamic applicability.}
	\label{fig:v2_dNdy_2relax}
\end{figure}
Figure~\ref{fig:v2_dNdy_2relax} shows the elliptic flow as a function of rapidity density for extreme values of the shear relaxation time factor. The data point labels are the centrality classes. The centrality class of 10\%--20\%, corresponding to the maximum magnitude, is apparent. The trend of the data points is similar to the previous analysis in~\citep{MyHydroStudy1}. The mismatch between the two elliptic flow curves begins at the centrality class of 55\%--65\%, and the abnormal increase in the flow beyond this point is also apparent. This beginning of elliptic-flow sensitivity to the extreme shear relaxation time is a signature of hydrodynamic onset~\citep{Romatschke2017Jan}, marked by the approximate yellow patch in figure~\ref{fig:v2_dNdy_2relax}.

\section{\label{sec:conclusion} Conclusion and outlook}
In an attempt to estimate the multiplicity scale below which hydrodynamics would become increasingly sensitive to the non-hydrodynamic mode for peripheral p-Pb collisions, the shear relaxation time of the fluid dynamics was varied over its extreme allowed range, while keeping the rest of the transport coefficients fixed. We utilized a recent state-of-the-art framework, JETSCAPE, to simulate the dynamics for each of the stages of the collision. We obtained elliptic flow results for two extreme shear relaxation times to study the onset.
A separation in the elliptic-flow curves in figure~\ref{fig:v2_dNdy_2relax} suggests that the applicability of hydrodynamics for centralities higher than 55\%--65\% is questionable, and that conservatively, the onset lies within a broad transition region, corresponding approximately to a rapidity density of $dN/dy \sim 10\text{--}15$, as marked by the yellow patch in figure~\ref{fig:v2_dNdy_2relax}. Whether there exists a sharp onset or a smooth transition region remains an open question. 

This range is consistent with our earlier analysis of peripheral Pb--Pb collisions at $\sqrt{s_{NN}} = 2.76$ TeV and Au--Au collisions at $\sqrt{s_{NN}} = 200$ GeV~\citep{MyHydroStudy1}, where the same procedure yielded an onset of $dN/dy \approx 10$. The present study extends this test to a p-Pb system, which differs from Pb--Pb and Au--Au not only in the size of the collision system but also in its geometry, being smaller and considerably more asymmetric. Finding a comparable onset scale despite these differences suggests that the multiplicity density at midrapidity, rather than the size or geometry of the colliding system, may be the more relevant control parameter for the breakdown of hydrodynamic behavior.
Both of these studies are suggestive of the possibility that a common region of multiplicity scale might lie between the initial out-of-equilibrium stage and hydrodynamization, across system sizes and collision energies.
Further investigation is therefore required in this direction, with additional systems/energies/model variations, to test this claim comprehensively.

It has been pointed out in~\citep{Spalinski2016} that the non-hydrodynamic mode may not decay during the relatively short lifetime of small-system collisions, in which case the non-hydrodynamic mode acts as the regulator of the theory. Therefore, it should be emphasized that the onset obtained in the present study is a conservative estimate, and the real onset may lie at lower values of $dN/dy$.  

The assumption that the soft sector can be investigated in isolation from the hard sector in high-energy collisions was more of a convenient approximation made in phenomenological studies, including the present one. However, a considerable amount of work has been done in the last few years toward a framework that can model a jet interacting with a hydrodynamic medium~\citep{Lokhtin2009, EPOS3_2014, Yan2018Mar,Chen2018Feb, Chen2020Nov, Chen2021Aug, Zhao2022Jan}. A recent review on this topic can be found in~\citep{Cao2023Sep}. The JETSCAPE framework has been designed to model the hard and soft sectors simultaneously, including the interaction between them. One of the crucial QGP signatures that we have not yet explicitly discovered in small systems is jet quenching~\citep{Nagle2018Oct_smallsys1}. A natural extension of this study would be to include a signature of the onset from the hard sector. It will be interesting to see how the inclusion of observables from the hard sector affects the soft-sector observables, and vice versa.

There is significant scope for improving this analysis. The Freestreaming time was held constant in this study. This value can be varied over a range, which could provide a quantifiable estimate of the uncertainty in the onset. The work can be performed with realistic 3+1D hydrodynamics such that the experimental flow at forward rapidity can be utilized for better tuning. The Bayesian constraint on $\eta/s$ with a parametrization as a function of temperature is another source of uncertainty that can be looked at in future studies.


\begin{acknowledgements}
N~H is grateful to Paul Romatschke, Chun Shen, Yasuki Tachibana, and Scott Moreland for clarifying various doubts from time to time. S~D acknowledges the SERB Power Fellowship, SPF/2022/000014, for support in this work. N~H acknowledges the use of Claude (Anthropic) for text proofreading and troubleshooting.

\end{acknowledgements}

\bibliographystyle{iopart-num}
\bibliography{hydro_study2.bib}
\end{document}